\documentstyle[epsfig,12pt]{article}
\begin{document}
\newcommand \be  {\begin{equation}}
\newcommand \bea {\begin{eqnarray} \nonumber }
\newcommand \ee  {\end{equation}}
\newcommand \eea {\end{eqnarray}}
\newcommand{\subs}[1]{{\mbox{\scriptsize #1}}}
\newcommand{\bm}[1]{{\mbox{\boldmath $#1$}}}
\title{{\bf MISSING INFORMATION AND ASSET ALLOCATION}}

\author{Jean-Philippe Bouchaud$^{1,2}$, Marc Potters$^2$\\ and Jean-Pierre
Aguilar$^2$}

\date{{\it\small
$^1$ Service de Physique de l'\'Etat Condens\'e,
 Centre d'\'etudes de Saclay, \\ Orme des Merisiers, 
91191 Gif-sur-Yvette Cedex, France \\ 
$^2$ Science \& Finance, 109-111 rue Victor-Hugo, 92532 France\\}
\vspace{1cm}June 1997 }

\thispagestyle{empty}
\maketitle

\begin{abstract}
When the available statistical information is imperfect, it is
dangerous to follow standard optimisation procedures to construct an
optimal portfolio, which usually leads to a strong concentration of
the weights on very few assets. We propose a new way, based on
generalised entropies, to ensure a minimal degree of diversification.
\end{abstract}
\newpage
One of the major problem in portfolio optimisation and diversification
is that the optimal solutions (for example in the risk/return sense of
Markowitz \cite{Mark}) are very often concentrated on a few assets, in
contradiction with the very idea of diversification. This feature is
considered as unreasonable by most practioners, who feel that it is
unwise to discard assets based on the fact that their past performance
(or their anticipated performance) is somewhat weaker than those
spotted out by the optimisation. Furthermore, due to the instability
of the covariance matrix and average return in time, the few assets
retained in the allocation tend to evolve with time; a situation which
is not satisfactory from an intellectual viewpoint, and furthermore
induces large costs. A way out is to impose to the optimisation
procedure to remain within preassigned upper and lower bounds for each
asset; however, this corresponds, to a large degree, to choosing one's
target portfolio, since many assets will have weights equal to the
minimum allowed weight.

The aim of the present paper is to introduce a family of
`diversification indicators' which measures how concentrated is the
allocation. The optimisation can be performed with a constraint which
precludes a too strong `localisation' of the weights on a few
assets. Perhaps more importantly, we relate these indicators to the
information content of the portfolio, in the sense of information
theory. A strongly concentrated portfolio corresponds to a large
information content, while equal weights to all assets corresponds to
a minimal information content. We argue that the rationale behind
trying to keep a balanced portfolio is that its information content
cannot be larger than the information available to the fund manager.
This intuitively makes sense: suppose that the average returns and
covariances are perfectly known and that one accepts the risk/return
framework. The choice of an optimal portfolio on the efficient border
is then totally justified, even if the weight of a single asset was
100\%. The problem is that one usually has only partial information:
past data have a finite length, therefore the determination of
statistical parameters is imperfect; predictions of future
volatilities and returns are obviously not entirely trustworthy --
this is why the optimal portfolio must reflect this lack of
information and keep a minimal degree of diversification. This can be
implemented using the idea of a `free-utility' function, in analogy
with the free-energy in thermodynamics.

Let $p_i$, $i=1,...,M$ be the weights of asset $i$ in a portfolio,
such that $\sum_i p_i=1$. The problem of characterising how uniform
(or how concentrated) these $p_i$ are has been investigated in a very
different context, namely that of the phase space of disordered
systems \cite{MP}. The following quantities were studied: \be Y_q =
\sum_i p_i^q \ee 
Obviously, for $q=1$, $Y_1=1$. Suppose now that all
weights are equal to $1/M$, where $M$ is the total number of
assets. Then $Y_2=1/M$ and thus goes to zero for large $M$. On the
other hand, if one of the asset has weight $p$ and all the others
weight $(1-p)/(M-1)$, one finds $Y_2=p^2 + O(1/M)$, which thus remains
finite for large $M$. $Y_2$ (and similarly for all $Y_q$, $q>1$) thus
distinguishes `localised' situations, where one, or a few assets
gather all the weight, from `delocalised' configurations,
corresponding to a high degree of diversification. Actually, $Y_2$ is
nothing but the {\it average weight}\/ of each asset.  We can thus
define an effective number of assets in the portfolio as
$M_\subs{eff}=1/Y_2$, and look for optimal portfolios with
$M_\subs{eff}$ greater or equal than some minimal value $M_0$. Let us
sketch the calculations within a classical Markowitz scheme. 

\begin{figure}
\centerline{\hbox{
\epsfig{file=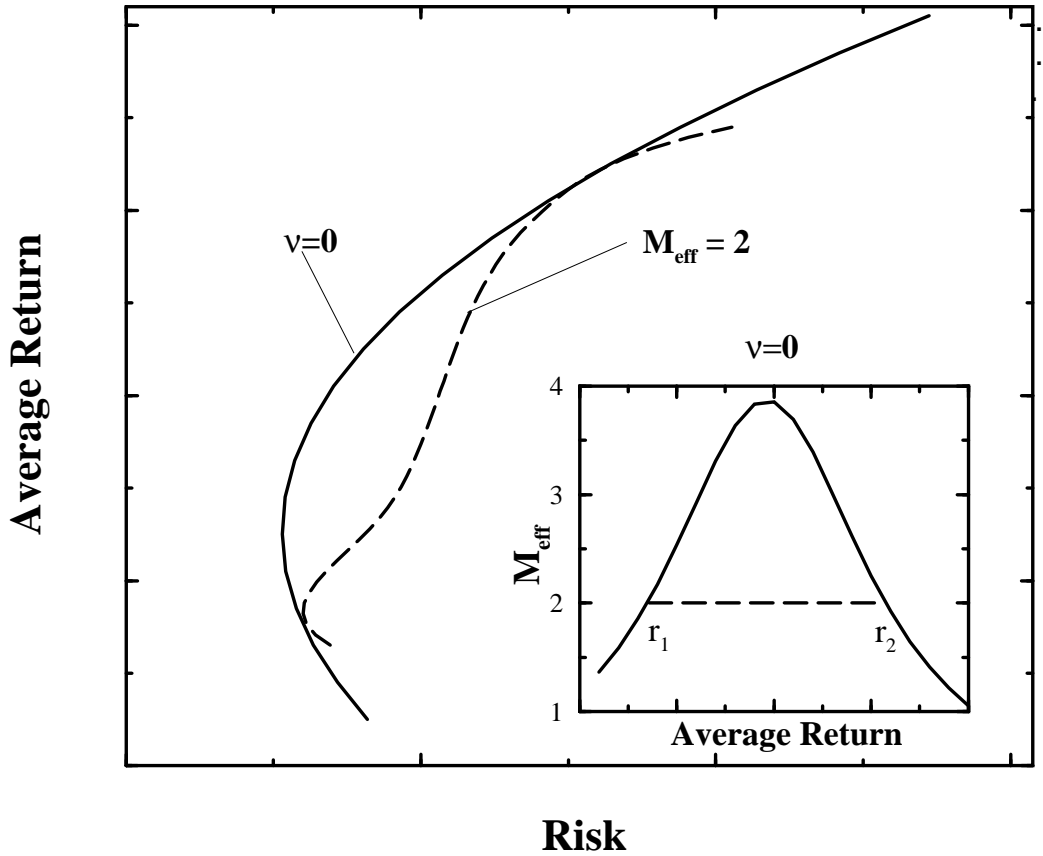,width=16cm}
}}
\caption{
Example of a standard efficient border $\nu =0$ (thick line) with four risky assets. If one imposes
that the effective number of assets is equal to $2$,
one finds the sub-efficient border drawn in dotted line. The inset shows the effective asset number
of the unconstrained optimal portfolio ($\nu=0$) as a function of average return. The optimal portfolio
satisfying $M_\protect\subs{eff}\geq2$ is therefore given by the standard portfolio for returns between
$r_1$ and $r_2$ and by the $M_\protect\subs{eff}=2$ portfolio otherwise.
} \label{Fig1} 
\end{figure}

One first checks whether the usual Markowitz solution satisfies the
minimum $M_\subs{eff}$ constraint; if not, one then looks for solution
constrained to $M_\subs{eff}=M_0$.  Introducing the covariance matrix
$C_{ij}$ and the expected return vector $r_i$, minimisation of the
variance of the portfolio for a given average return and for a fixed
effective number of assets leads to:
\be
\frac{\partial }{\partial p_i}\left[-\sum_{jk} p_j C_{jk} p_k + \lambda \sum_j p_j 
r_j + \mu \sum_j p_j - \nu \sum_j p_j^2 \right] =0 \label{opt}
\ee
or, in matrix notation:
\be
\bm{p} = \frac{1}{2} {\cal C}^{-1} [\lambda \bm{r} + \mu\bm{1}]\label{mm}
\ee
with ${\cal C}_{ij}=C_{ij}+\nu \delta_{ij}$.  The three Lagrange
parameters $\lambda,\mu,\nu$ are determined as to satisfy the three
constraints (sum of weights normalised to 1, fixed average return $R$
and fixed $M_\subs{eff}= 1/Y_2$). This generates a family of
`sub-efficient border' which is drawn in \ref{Fig1}, together with the
usual efficient border, which corresponds to $\nu=0$ (no constraint
on $Y_2$).  Technically, Eq. \ref{mm} requires, as in the usual
case, the inversion of a matrix which is the covariance matrix plus a
constant along the diagonal; this constant must be adjusted to satisfy 
the constraint. Optimisation with a fixed $Y_q$ with $q
\neq 2$ is of course possible, although the calculations are somewhat
more complex.

The relation with information theory  is based on the fact that the indicators
$Y_q$ are actually generalised entropies \cite{tsallis}. Indeed, the classical definition of
the entropy (or missing information) associated to the weights $p_i$ is \cite{Shannon}:
\be
{\cal S} = - \sum_i p_i \log p_i
\ee
which is zero if the portfolio is concentrated on one asset, and maximum for an
uniform portfolio ($p_i=1/M$). It is easily checked that $\cal S$ can be
expressed in terms of the $Y_q$'s as:
\be
{\cal S} = - \lim_{q \to 1} \frac{Y_q-1}{q-1}
\ee
All the $Y_q$'s play a role similar to the entropy, and can be used to
limit the intrinsic informational content associated to the very
choice of the weights: the larger the available information, the
larger the allowed values of $Y_q$. It is clear that in the absence of
information, the only rational choice is $p_i=1/N$; while complete
information corresponds to $\nu=0$. It is interesting to note an
analogy with thermodynamics: at zero temperature (absence of
uncertainty), one must minimise the energy, which is the analogue of
(minus) a utility function. At nonzero temperature, one must minimise
a `free-energy' which has both an energetic and entropic
contribution. Similarly, one can introduce a general `free-utility'
${\cal F}_q$ which contains an entropic contribution:
\be
{\cal F}_q = U - \nu \frac{Y_q-1}{q-1}
\ee
where $U$ is a utility function. The above optimisation, Eq. \ref{opt}, corresponds to
$U = \lambda R - \sigma^2$, where $\sigma^2$ is the variance of the portfolio. $\nu$
plays the role of a `temperature', which is high when the uncertainty is large. It is
obvious that if $\lambda=0$ (no constraint on the average return), one finds $p_i=1/M$
in the limit where $\nu \to \infty$. Note that the dimension of $\nu$ is the same as that of $\sigma^2$. It could be interesting to work in a `canonical'
ensemble where $\nu$, rather than $Y_2$, is fixed. A natural choice for $\nu$ 
would then be the squared error in the average returns $r_i$.

In summary, we have argued that optimisation under incomplete
information should be performed with constraints on generalised
entropies (which are actually diversification indicators), much like
in standard thermodynamics. This generates new sub-efficient borders,
and allows one to avoid undesired overconcentration of optimal
portfolios one very few assets, in accord with general market
practice. This idea is not restricted to the Markowitz optimisation
scheme, and is useful also when the risk is measured, not as a
variance, but as the Value-at-Risk, which in a non-Gaussian world
leads to different optimal portfolios \cite{us}.

\clearpage

\end{document}